\documentclass[pre,twocolumn,superscriptaddress]{revtex4}
  \usepackage{graphicx}
  \usepackage{pslatex}

\begin{document}

 \title{Bi-clique Communities}
 \author{Sune Lehmann}
   \affiliation{Center for Complex Network Research and Department of Physics, Northeastern University, Boston, MA 02115, USA}
   \affiliation{Center for Cancer Systems Biology, Dana-Farber Cancer Institute, Harvard University, Boston, MA 02115, USA}
   \affiliation{Informatics and Mathematical Modelling. Technical University of Denmark. DK-2800 Kgs.~Lyngby, Denmark}
 \author{Martin Schwartz}
   \affiliation{Informatics and Mathematical Modelling. Technical University of Denmark. DK-2800 Kgs.~Lyngby, Denmark}
   \affiliation{IT University of Copenhagen, DK-2300 Copenhagen S, Denmark.}
 \author{Lars Kai Hansen}
   \affiliation{Informatics and Mathematical Modelling. Technical University of Denmark. DK-2800 Kgs.~Lyngby, Denmark}

\date{\today}

\begin{abstract}
We present a novel method for detecting communities in bipartite networks. Based on an extension of the $k$-clique community detection algorithm, we demonstrate how modular structure in bipartite networks presents itself as overlapping bicliques. If bipartite information is available, the bi-clique community detection algorithm retains all of the advantages of the $k$-clique algorithm, but avoids discarding important structural information when performing a one-mode projection of the network. Further, the bi-clique community detection algorithm  provides a new level of flexibility by incorporating independent clique thresholds for each of the non-overlapping node sets in the bipartite network.  
 \end{abstract}

 \maketitle

\section{Introduction}
The theoretical understanding of the structure and function of complex networks has grown rapidly during the past few years \cite{dorogovtsev:02,newman:06uq,caldarelli:07fk}. One large component of the field of complex networks regards the study of community structure in networks; for reviews see \cite{newman:04d,danon:05a}. Community structure describes the property of many networks that nodes are divided into `communities' with many intra-community links and sparse connections between the densely connected modules. In spite of a focused research effort, the mathematical tools developed to describe the structure of large complex networks are continuously being refined and redefined.

Currently, the endeavour of detecting community structure in complex networks can be divided into two main approaches. One main class can be labeled \emph{global methods}, of which the most notable example is the modularity introduced by Newman and Girvan \cite{newman:04a}; global methods regard community detection as a global optimization problem, where the objective function is particular to each method. Due to the complexity of such optimization problems, the global methods are typically stochastic in nature. The other class is \emph{local methods}, where the best known example is the $k$-clique method described by Palla \emph{et al.}~\cite{palla:05a,palla:07kx}; here, local structural information is utilized to reveal the community structure of a network. The local methods are usually deterministic.

Although widely studied in the fields of statistics and computer science \cite{dhillon:01ly,sinkkonen:03vn,ding:06ys,reiss:06zr}, the study of bipartite networks and their community structures has only recently been moving into the focus of the network community. So far, all efforts have been focused on global community detection methods \cite{guimera:07fk,barber:07uq,zhang:07kx}. Here we present a simple algorithm---based on a local framework---that has considerable power, flexibility, and accuracy.

\section{Bipartite networks}
A bipartite network is a network with two non-overlapping sets of nodes $\Delta$ and $\Gamma$, where all links must have one end node belonging to each set. As is clear from the examples below, many real world networks are naturally bipartite:

\noindent $\bullet$ \emph{Social Networks}. The available data regarding many different social networks consist of what is known as `affiliation networks'. Examples of affiliation networks include the scientific collaboration network \cite{newman:01a,newman:01b,guimera:05vn} (where the two node sets consist of papers and authors, respectively), the movie-actor network, where the network edges connect an actors and films \cite{imdb}, and artistic collaboration networks \cite{guimera:05vn}, where a link indicates the participation of a creative team. Other examples of social networks that can be inferred from bipartite data are the movie-recommendation network \cite{netflix} that links users to the movies they have watched, or the song-listener network that link music listeners to the music they play on their computer \cite{lastfm,lambiotte:05fk}.

\noindent $\bullet$ \emph{Biological Networks}. Many important types of biological networks are naturally bipartite. 
Examples of bipartite biological networks are the metabolic network, where the two types of nodes are reactions and metabolites \cite{jeong:00vn}, the human disease network of genes and diseases \cite{goh:07ys}, and the network describing drugs and their molecular targets \cite{yildirim:07zr}.

\noindent $\bullet$ \emph{Information Networks}. The bipartite structure is also very common for information networks. The generic example is a word-document network, where one type of nodes is documents (web-pages, emails, dictionary entries, etc) that link to the words they contain \cite{wordnet,hofmann:98uq,hofmann:98kx,hofmann:99fk}

Most of the studies of real world networks listed above, do not analyze the bipartite networks directly, but rather one-mode projections of the network. Below, we will demonstrate how the one-mode projection of a bipartite network disregards important network information and argue that a direct analysis of the bipartite network is a more natural option that captures important nuances of the network structure that are invisible to the analyses based on unipartite projections.

A bipartite network has a bipartite $(n_\Delta \times n_\Gamma)$ adjacency matrix $E$, where $n_\Delta$ and $n_\Gamma$ are the number of nodes in each set. This matrix is constructed such that
\begin{equation}
E_{ij} = \left\{
\begin{array}{ll}
	1 & \text{if there is a link between node  $i$ and $j$, and}\\
	0 & \mathrm{otherwise.}
\end{array}
\right.
\end{equation}
In real networks, this matrix is typically very sparse. Any bipartite network can be transformed into two unipartite networks. One network consisting of the $n_\Delta$ nodes in the $\Delta$ set and one network consisting of the $n_\Gamma$ nodes in the $\Gamma$ set. These one-mode projections are obtained by calculating the two symmetric, weighted matrices the $A_\Delta = E E^T$ and $A_\Gamma = E^T E$. The diagonal elements $A_{ii}$ of these matrices contain the number of links connected to node $i$ in the bipartite network, and the off-diagonal elements $A_{ij}$ contain information on the number of nodes in the complementary set are shared by nodes $i$ and $j$.

The conceptual simplicity of the one-mode projection comes at a high cost. First of all, the procedure typically eradicates the sparsity of the $E$ matrix; this is especially problematic, when constructing the adjacency matrix for the smaller set of nodes, in the case where one of the node sets is significantly larger than the other. Secondly, much of the information present in the bipartite state becomes encoded in the weights of the adjacency matrix. However, due to (1) technical difficulties regarding the analysis of weighted matrices\footnote{In fact, most community detection methods assume that networks are unweighted and undirected. See for example \cite{palla:05a,newman:06a}.} and (2) the high link-density of the one-mode projections (if the adjacency matrix is densely populated, all nodes are connected and the network has very little structure), these matrices are usually thresholded such that only entries higher than some threshold are retained. Similarly, the diagonal of the one-mode adjacency matrices is usually set to zero, since self-links are not of interest in the subsequent network analysis.

\begin{figure}
\centering
\begin{tabular}{ccc}
\includegraphics[width=.295\hsize]{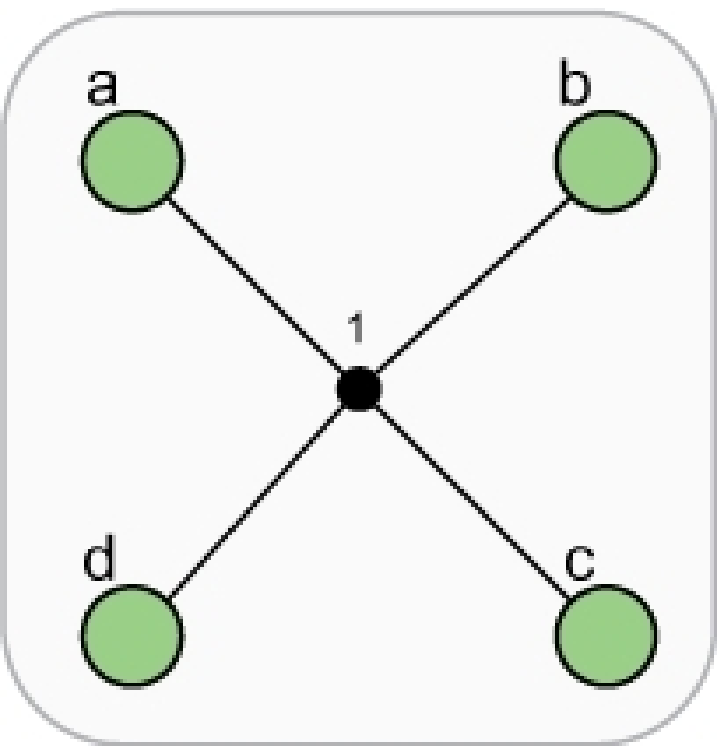} & \includegraphics[width=.295\hsize]{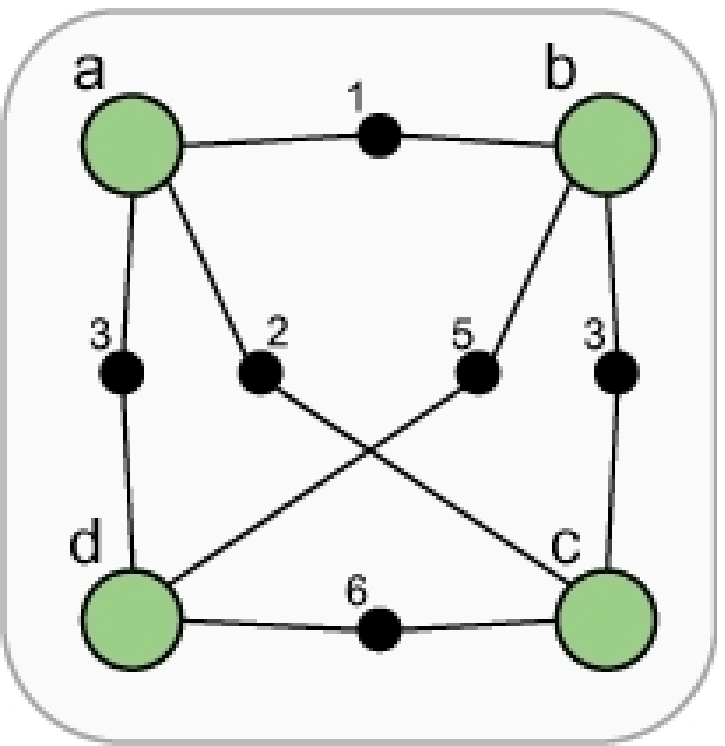} & \includegraphics[width=.295\hsize]{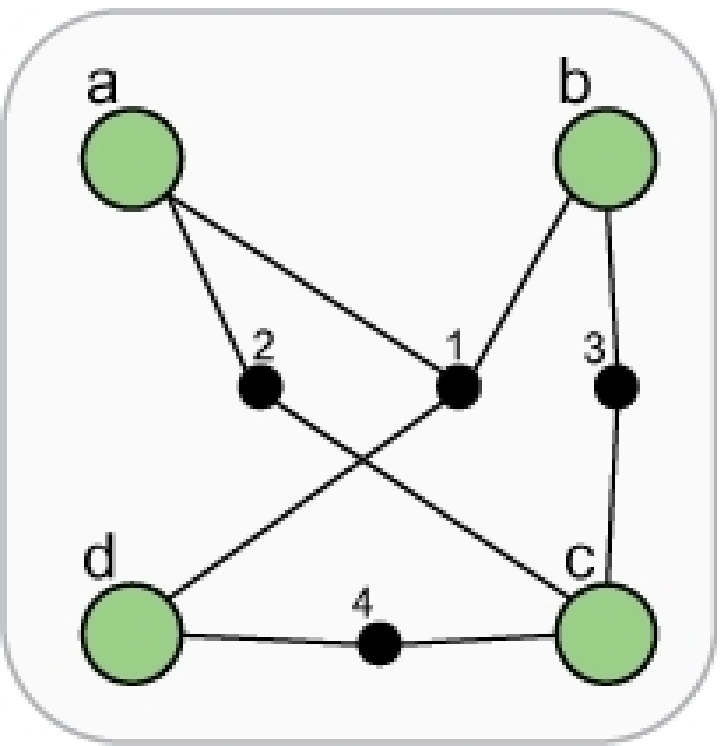}\\
(a) & (b) & (c)
\end{tabular}
\caption{Color online: Three distinct bipartite networks that result in identical one-mode projections. In the first case, (a), the nodes $\Delta = \{a,b,c,d\}$ share the single node $\Gamma = \{1\}$ in the complementary set. (b) The second case, that includes the nodes $\Delta = \{a,b,c,d\}$ and the complementary nodes $\Gamma = \{1,2,3,4,5,6\}$, has every pair of nodes in the $\Delta$ level linked via different nodes in the complementary set. In the third case, (c) three of the the four $\Delta = \{a,b,c,d\}$ nodes, $(a,b,d)$, share a single node in the complementary node set, while all other linkages between $\Delta$-nodes in this network are pair-wise and run via nodes in the complementary set that are exclusive to the two nodes linked.
}\label{fig:projectionproblem}
\end{figure}

One aspect that is rarely discussed in the literature is the fact that even if we keep all the off-diagonal weights in the one-mode adjacency matrix, essential information is lost when performing the one mode projection. This is clear from the fact that we cannot reconstruct $E$ from $A_\Delta$ and $A_\Gamma$. It is, however, instructive to study precisely what information is lost. Specifically, the problem is that the one-mode adjacency matrices only contain two-point correlations. Given two nodes, $i$ and $j$ in one of the sets, the corresponding adjacency matrix informs us how many nodes these two share in the complementary set. Given a third node $k$, we also know the number of nodes that are shared by $i$ and $k$ or $j$ and $k$, respectively in the complementary set, but we have no information about which nodes from the complementary set that $i$, $j$, and $k$ connect to in common: The same set of nodes could be shared by $i,j$, and $k$, or the nodes in the complementary set could be shared pairwise, but not among all three. A practical example of this problem is shown in Figure~\ref{fig:projectionproblem}.

In Figure~\ref{fig:projectionproblem} we display 3 simple bipartite networks. The network described in Figure~\ref{fig:projectionproblem}~(a) shows a case where all $\Delta$ nodes are linked to a single node in the $\Gamma$ set. A practical example of this motif can be found in the movie-actor network, where this would be the case when 4 individuals act together in a single film. In Figure~\ref{fig:projectionproblem}~(b) a different network is displayed. Here, all four nodes in the $\Delta$ set are interconnected via pair-wise links to six distinct nodes in the $\Gamma$ set. In the movie-actor network this corresponds to four actors who have all been in films together, but with precisely two common actors per film; \emph{these six movies could be far apart in time and space}. Therefore the significance of this network motif is very different from the significance of the motif displayed in Figure~\ref{fig:projectionproblem}~(a).  Finally, the network in Figure~\ref{fig:projectionproblem}~(c) lies somewhere in between the two other cases.

Important qualifying information about the nodes shared in the complementary set is not carried over in the one mode projection of the network. When we perform the one-mode projection of each of these three networks onto the $\Delta$ nodes (we retain the weights but remove the diagonals), the one-mode adjacency matrices become
\begin{equation}
A_{(a)}=A_{(b)}=A_{(c)}=\left(
\begin{array}{cccc}
0 & 1 & 1 & 1\\
1 & 0 & 1 & 1\\
1 & 1 & 0 & 1\\
1 & 1 & 1 & 0
\end{array}
\right).
\end{equation}
In the one-mode projection the three networks become indistinguishable $4$-cliques.

In summary, the one-mode projection approach disregards important network information in two distinct steps. Firstly, when the projection itself is performed, all (sparse) information about the bipartite linkages is reduced to a dense network of two-point correlations. Secondly, all of the information contained in the weights is typically discarded in a subsequent thresholding operation. In the following section, we will explain a simple way of analyzing the community structure of the bipartite network directly.

\section{Biclique communities}
\begin{figure}
 \centering
 \includegraphics[width=.9\hsize]{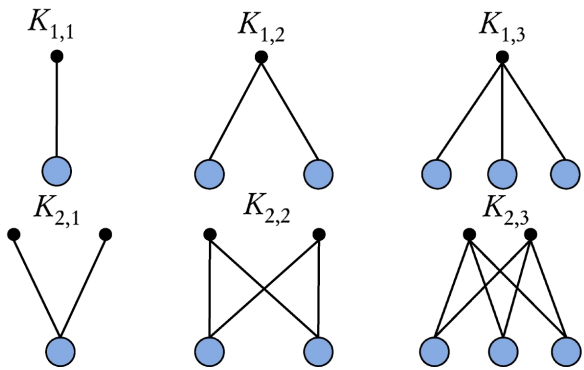}
\caption{Color online: Maximally connected bigraphs. The notation $K_{a,b}$ means that the complete bigraph consists of $a$ of the black nodes in the $\Delta$ set and $b$ of the larger nodes nodes in the $\Gamma$ set.}\label{fig:completegraphs}
\end{figure}

\begin{figure}
 \centering
 \includegraphics[width=.8\hsize]{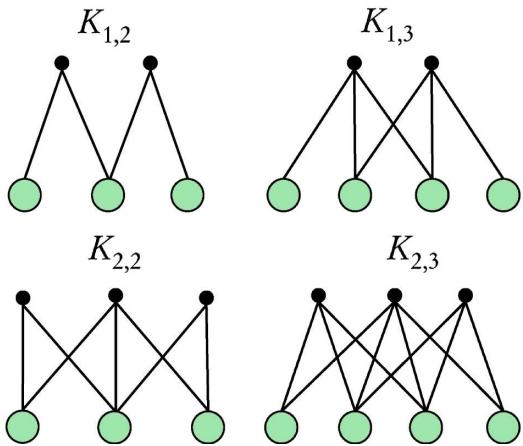}
\caption{Color online: Biclique adjacency. Two $K_{a,b}$ cliques are adjacent if they share at least a $K_{a-1,b-1}$ clique. In this figure we list a few examples. The two adjacent $K_{1,2}$ cliques share a $K_{0,1}$ biclique, the two adjacent $K_{1,3}$ cliques share a $K_{0,2}$ clique, the  two adjacent$K_{2,2}$ cliques overlap by a $K_{1,1}$ clique, and the two adjacent $K_{2,3}$ cliques share a $K_{1,2}$ clique.}\label{fig:bicliqueoverlap}
\end{figure}

In analogy with the unipartite case, the basic observation on which our community definition relies is that a typical community consists of several complete sub-bigraphs\footnote{In the following we will be discussing bipartite networks almost exclusively. For simplicity, we will drop the prefix `bi' and simply write `clique', when we mean `bi-clique', etc, when there is no danger of confusion.} that tend to share many of their nodes. A number of complete bipartite graphs are displayed in Figure~\ref{fig:completegraphs}.
We now define a $K_{a,b}$ clique as a complete subgraph with $a$ nodes in the $\Delta$ node set and $b$ nodes in the $\Gamma$ node set. A $K_{a,b}$ clique can be identical to a maximal complete subgraph or it can exist on a subset of the nodes of a maximal complete subgraph.
Generalizing from \cite{palla:05a}, we now define a $K_{a,b}$ clique community, as a union of all $K_{a,b}$ cliques that can be reached from each other through a series of \emph{adjacent} $K_{a,b}$ cliques. We define two $K_{a,b}$ cliques to be adjacent if their overlap is at least a $K_{a-1,b-1}$-biclique. Another way of saying this is that the two cliques must share at least $a-1$ upper vertices and $b-1$ lower vertices. See Figure~\ref{fig:bicliqueoverlap}.

An important feature of the biclique community approach is that the biclique method provides an immediate \emph{context} to the communities that are detected. In the movie-actor network, a list of actors is always accompanied by a list of film. It is immediately clear why the actors in a group belong together---we know the ouevre that they share. In the metabolic network a list of metabolites is accompanied a list of the reactions they participate in; this presence of context is an important help in determining the function of detected communities. In this sense, the bi-community information is more valuable that the one obtained by finding structure in the two unipartite projections because it provides specific links between the communities that are present in the two node sets; we will discuss precisely what we mean by this in the next section. The biclique method described here is a related to co-clustering,  \cite{dhillon:01ly,sinkkonen:03vn,ding:06ys,reiss:06zr}.

\section{Relation to $k$-clique communities}
When bipartite network information is available, the biclique community detection method is an attractive alternative to the $k$-clique algorithm. The $k$-clique algorithm is unable to analyze sparse network regions. This is due to the fact that $2$-clique communities are simply the connected components of the network and are contain little information about the network structure. The first non-trivial $k$-clique has size $k = 3$. These two facts combined, result in the inability to analyze sparse network regions---simply because nodes must have at least two links in order to qualify for participation in a $3$-clique. In networks with heavy tailed degree distributions, a large fraction of the nodes have degree less than two and an even larger fraction of nodes do not participate in cliques of size three or greater. 

If bipartite data is available, the biclique method is able to detect subtle structures. In order to understand why this is the case, it is useful to consider the relation between the two methods. We begin by revisiting Figure~\ref{fig:projectionproblem}. In terms of cliques, Figure~\ref{fig:projectionproblem}~(a) corresponds to a $K_{4,1}$ clique exemplified by four authors part of the same movie. Figure~\ref{fig:projectionproblem}~(b) corresponds to six adjacent $K_{2,1}$ cliques joined in a $K_{2,1}$ community. Finally, Figure~\ref{fig:projectionproblem}~(c) can be recognized as one $K_{3,1}$ clique and three $K_{2,1}$ cliques. When considering the community structure the small network in Figure~\ref{fig:projectionproblem}~(c), all nodes are included if we set the threshold at $K_{2,1}$, but we only include the nodes $\Delta = \{a,b,c\}$ and $\Gamma = \{1\} $ if we raise the threshold and look for $K_{3,1}$ communities. In this small example, we use the biclique technique to look 'inside' the $4$-clique that arises when we project the small bipartite networks onto the $\Delta$ nodes.

The biclique communities have clear translations in terms of the two un-thresholded one-mode projections. The $K_{2,1}$ communities correspond to connected components in the projection onto the $\Delta$ nodes; the two $\Gamma$ nodes in each of the cliques $K_{2,1}^{(1)}$ and $K_{2,1}^{(2)}$ are linked in the one-mode projection onto the set of $\Gamma$ nodes if the two cliques share a $K_{1,0}$ clique , that is, if the two cliques are adjacent. Similarly, the $\Gamma$ nodes in each of the two cliques $K_{2,1}^{(2)}$ and $K_{2,1}^{(3)}$ are also linked in the one-mode projecton onto the $\Gamma$ network if they share a $K_{1,0}$ clique. Thus the community of the three adjacent $K_{2,1}$ nodes corresponds to a connected set of nodes (a $2$-clique community) in the network of $\Gamma$ nodes. This small example is easily generalized to the case of $n$ adjacent $K_{2,1}$ cliques. A similar argument shows that $K_{1,2}$ communities correspond to connected components in the $\Gamma$ networks. What is particularly noteworthy here, is that from the bipartite community detection algorithm---in addition to the connected components---we also get a list of nodes in the complementary set of nodes that \emph{correspond} to the connected components. These nodes do not necessarily form a connected component in the complementary one-mode projection.

The result mentioned in the previous paragraph is readily generalized. In fact, $K_{a,1}$ and $K_{1,b}$ biclique communities correspond to $a$- and $b$-clique communities in the projections onto $\Delta$ and $\Gamma$ nodes, respectively. A clique $K_{a,1}^{(1)}$ results in a $a$-clique (consider Figure~\ref{fig:projectionproblem}~(a)); another clique $K_{a,1}^{(2)}$ results in another $a$-clique. Now, if these two share a $K_{0,a-1}$ clique (e.g in the movie-actor network, this would correspond to sharing $a-1$ actors), then these $a-1$ nodes are fully connected an therefore a $a-1$ clique. In other words, this corresponds to a $a$-clique community in the $\Delta$ one-mode projection. This result can be generalized to the case of $K_{1,b}$ biclique communities and $b$-clique communities in the $\Gamma$ projection. As is clearly illustrated in the examples displayed in Figure~\ref{fig:projectionproblem}, this result is \emph{not valid} going from the one mode projection to the bipartite case.

In general, the biclique communities have the following relationship to the one-mode projections: A $K_{a,b}$ community corresponds to
\begin{enumerate}
\item An $a$-clique community $D$ in the projection onto the $\Delta$ nodes.
\item A $b$-clique community $G$ in the projection onto the $\Gamma$ nodes.
\item Further, in order to qualify for membership in the community $D$, a node must connect to a node in $G$ and vice versa.
\end{enumerate}
This is precisely why the biclique algorithm presented here is able to detect structures between $2$-clique communities and $3$-clique communities where the $k$-clique algorithm fails to locate structure. The $K_{2,2}$ clique communities, for example, are simply connected components in each one-mode projection, with the \emph{additional constraint} that the connected component in each projection must be correlated with the complementary connected component as described in item 3. above. This is the precise content of the argument in the previous section that the biclique algorithm provides context to the communities. We emphasize that all of the arguments presented here apply to the un-thresholded version of the one-mode projections---thresholding the one-mode projections enhances the advantages of the biclique community detection method.

The $K_{a,b}$ clique community method possesses the advantages of the $k$-clique algorithm. The most important strength of the $k$-clique method is that distinct communities can \emph{overlap} by sharing their nodes. This ability is essential when analyzing many real networks: Consider social networks: In social networks, most actors participate many communities of family, friends, and work relations. The biclique algorithm presented here allows the same type of overlap---nodes in the $\Delta$ set can overlap with other $\Delta$ nodes and similarly for the $\Gamma$ set. Cases where there is overlap between nodes from both sets of nodes are particularly interesting. As it the case with the $k$-clique algorithm, the node overlap allows the user to zoom out and observe the network of communities, linked by common nodes.

Another well known advantage of the $k$-clique method is that it allows the user to change the resolution at which the network is observed, by adjusting the clique size $k$. A high value of $k$, allows the user to observe structures in the denser regions of the graph, whereas low values of $k$ allows the user to study the structure of the sparser regions of the network. In the case of the $K_{a,b}$ cliques, this ability is enhanced because we are able to vary the sizes of $a$ and $b$ independently of each other. As an example, consider the movie-actor network. We can search for groups of actors that have acted as ensembles by choosing $a$ to be low and $b$ to be high, or we search for a series of films share a small group of actors by choosing a high $a$ and a small $b$. By varying $a$ and $b$, we can systematically probe different aspects of the community formations by studying the size distributions of communities and by visual inspection \cite{palla:05a,bcfinder}. Section \ref{sec:noc} elaborates on this point.

\section{Detecting biclique communities}
The biclique communities are detected by a procedure analogous to the on presented for $k$-clique detection in \cite{palla:05a}, however, some of the steps in the detection algorithm are different. We will describe the algorithm for detecting communities of size $K_{a,b}$ in the following. 

\emph{Enumerate Maximal Bicliques:} To find the biclique communities, we begin isolating the $N$ maximal bicliques in the bipartite network under study. We use a freely available algorithm LCM (Linear time Closed itemset Miner) version 4.0 \cite{uno:05uq} (downloaded from \cite{lcm}) for this purpose.  Using the list of maximal bicliques, we construct two $(N \times N)$ symmetric clique-overlap matrices $L_\Delta$ and $L_\Gamma$. The matrix elements of $L_\Delta$ contain information about the clique overlap among the nodes in the $\Delta$ set. Along the diagonal, this matrix contains the number of $\Delta$-nodes in maximal biclique $i$. The off-diagonal matrix elements contain the number of $\Delta$ nodes that maximal biclique $i$ and maximal biclique $j$ have in common. The matrix $L_\Gamma$ is similar but describes the overlap amongst the $\Gamma$-nodes.

\emph{Threshold overlap matrices:} The thresholding procedure goes through several steps. The first step evaluates the diagonal elements. Diagonal elements greater than or equal to $a$ are set to one, all other diagonal elements are set to zero. We then threshold the off-diagonal elements; this step is slightly more involved than the corresponding step in the $k$-clique algorithm. First we set all elements of columns and rows that correspond to a zero diagonal element to zero. Next, we threshold the remaining elements, keeping only elements greater than or equal to $a-1$. We carry out the same procedure for matrix $L_\Gamma$, using $b$ in the place of $a$ in the instructions above. Each of the thresholded overlap matrices (let us call them $L_\Delta'$ and $L_\Gamma'$) now contain information about the overlap in each of the two sets of nodes. In order for us to find the $K_{a,b}$ clique community information, we now create the final total overlap matrix $L$ by only accepting the clique overlap, when it is present in both of the individual matrices, so we set $L = L_\Delta' \wedge L_\Gamma'$, where $\wedge$ is the logical operation AND. The total clique overlap matrix, $L$, informs us about what maximal cliques are \emph{adjacent} in the $K_{a-,b-1}$ sense. 

\emph{Find connected components:} The final step is to determine the connected components of $L$; each component corresponds to a biclique community. From the maximal bicliques that are members of each community, we extract the indices of nodes that participate in each biclique community.
\begin{figure}
\centering
\begin{tabular}{c}
\includegraphics[width=\hsize]{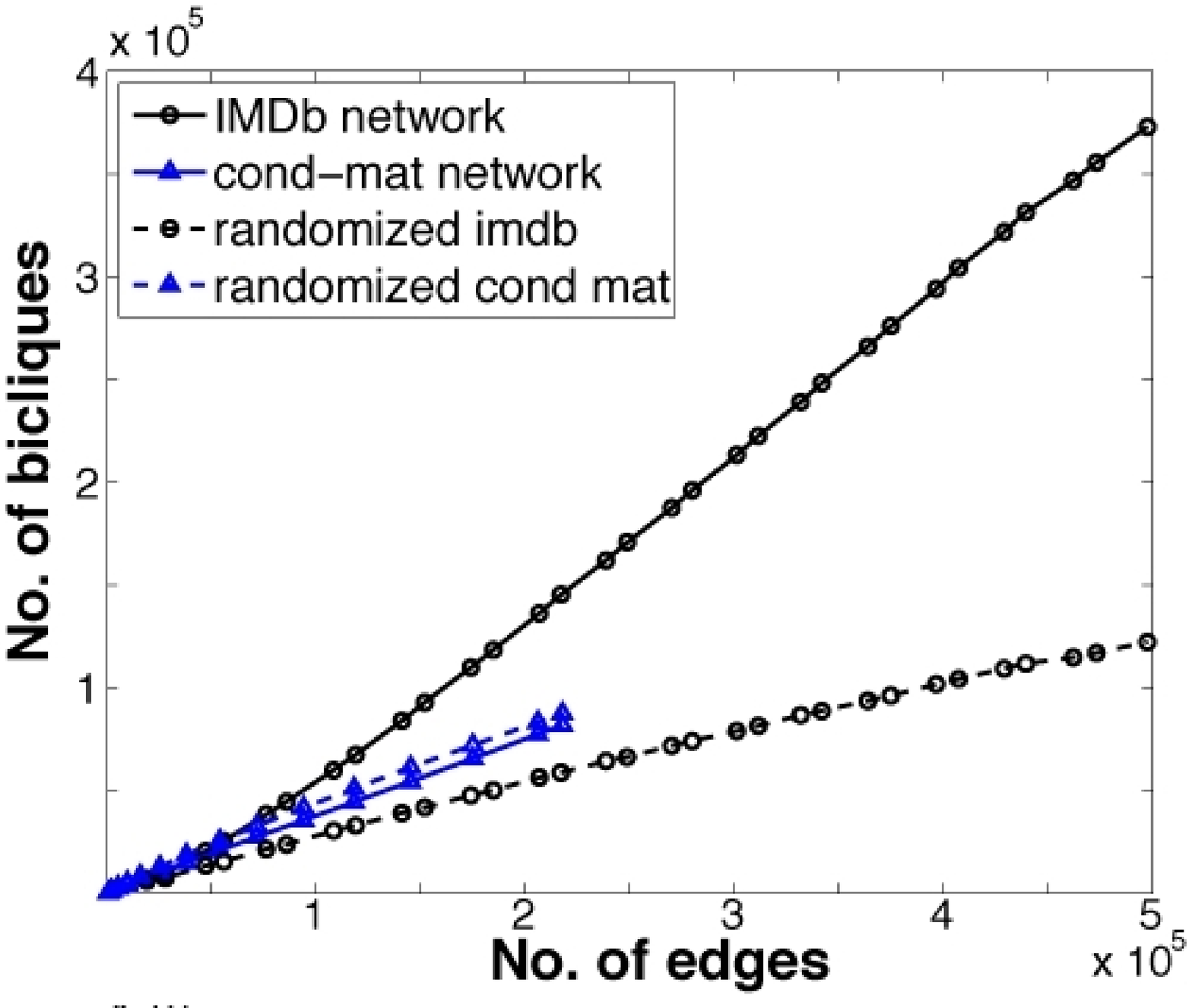} \\ 
(a) \\
\includegraphics[width=\hsize]{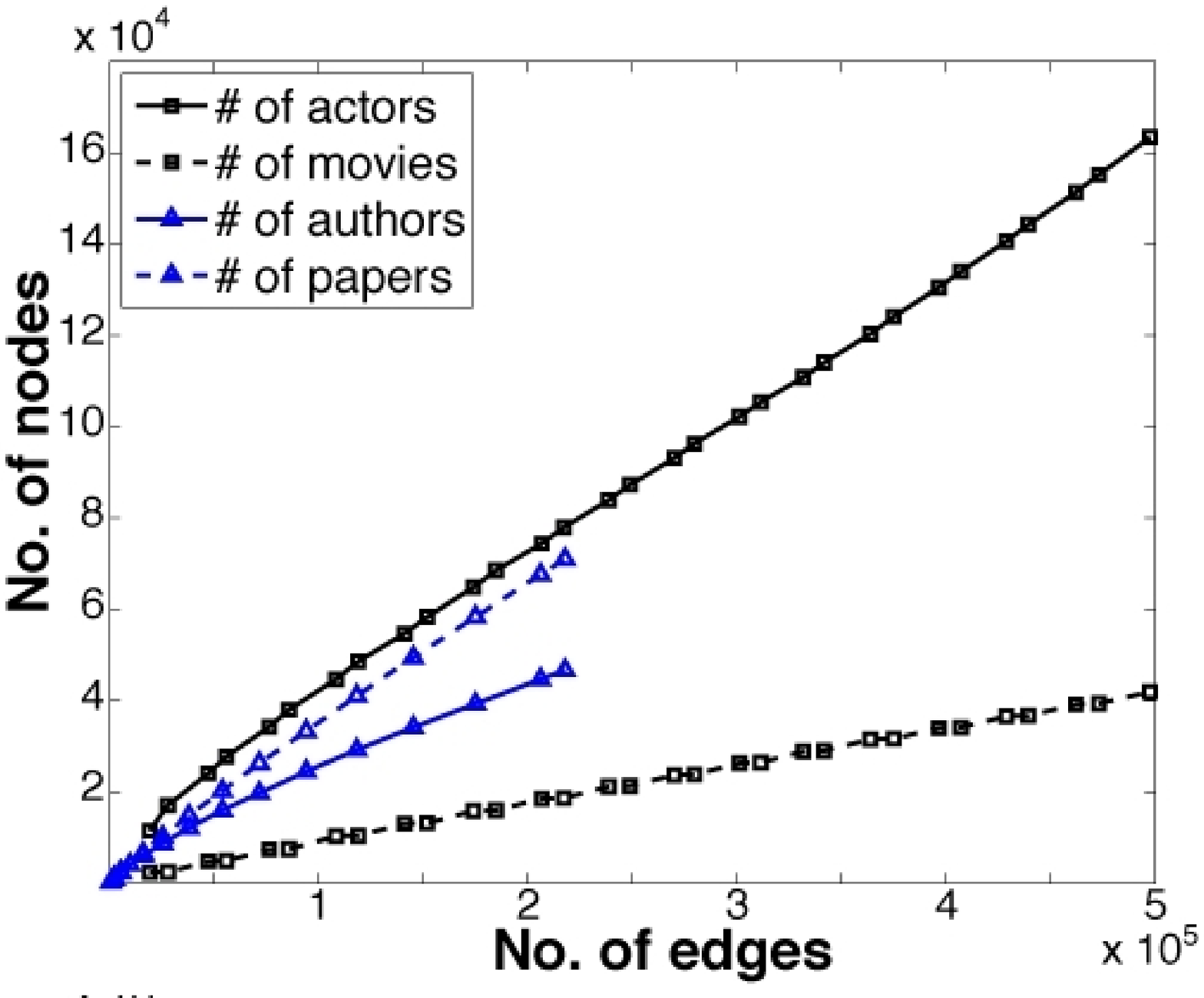} \\
(b)
\end{tabular}
\caption{Color online: In many sparse real world networks, the number of maximal bicliques grows linearly with the size of the input data base. Panel (a) shows the number of maximal bicliques found for the IMDb \cite{imdb} and cond-mat \cite{arxiv} networks  as a function of size of the networks (measured in number of edges). The solid line labeled by circles shows is the number of bicliques found in the IMDb data and the dashed line labeled by circles is the number of bicliques in the randomized version of the same network; the lines labeled by triangles show the same quantities for the cond-mat network. The network was randomized using a bipartite version of the algorithm suggested in \cite{maslov:02}. There are significant differences between the real and randomized data sets in the IMDb data, whereas there is little change for the cond-mat data. These differences are mainly due to the fact that, on average, there are more actors involved in the production of movies than there are authors of scientific papers. A forthcoming paper discusses the subject of biclique motifs in various bipartite networks.  Panel (b) shows the growth of the bipartite adjacency matrix as a function of the number of edges included in the analysis; solid line marked by squares is the number of distinct movies and the dashed line marked by squares is the number of actors; the lines labeled by triangles display the number of authors (solid line) and the number of papers (dashed line) for the cond-mat network. The incremental growth of the number of movies in the IMDb network is explained in the main text. 
}\label{fig:complexity}
\end{figure}

\section{Network of Communities}\label{sec:noc}
\begin{figure*}
\includegraphics[width=\hsize]{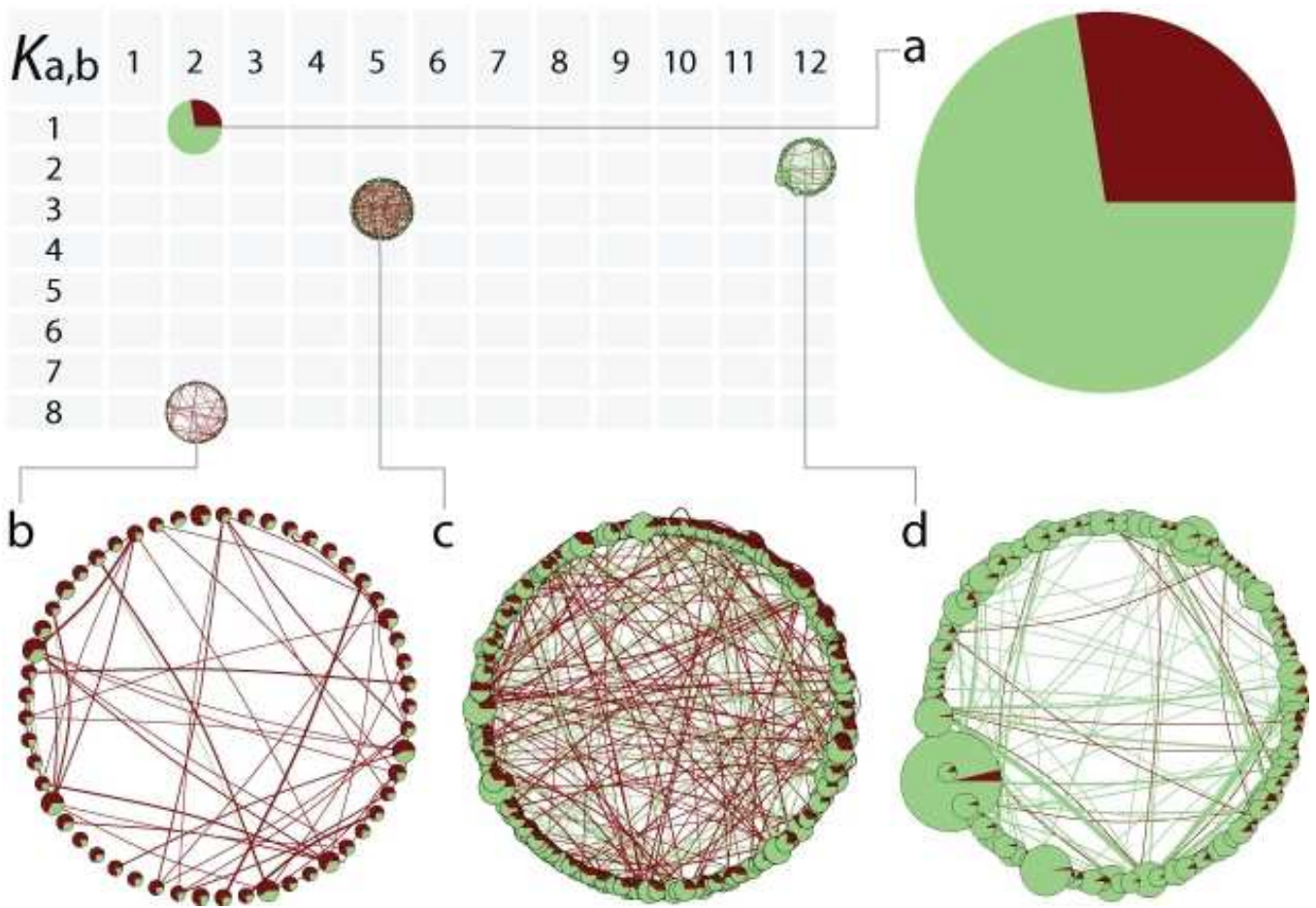} 
\caption{Color online: Networks of communities for in cond-mat network \cite{arxiv} for various choices of $K_{a,b}$. In these plots, authors are represented by the dark red and papers are represented by light green; thus author node-overlap is shown as a dark red link and paper overlap is shown as a light green link. Panel~\textbf{a} shows the network of communities for $K_{1,2}$, panel~(b) shows the network of communities for $K_{8,2}$, panel~(c) shows the network for $K_{3,5}$ and panel~(d) describes the case of $K_{2,12}$. See the main text for details. All panels are screen-shots from BCFinder \cite{bcfinder} \label{fig:communitynet}}
\end{figure*}

It is possible to construct a network consisting of the biclique communities. In this network, each community is a node and two communities are linked if they have nodes in common. Nodes from each partition of the network are allowed to overlap, so the network has two types of links ($\Delta$-links and $\Gamma$-links), the number of overlapping nodes can be encoded as the link-weight. Since the communities have different sizes and we would like to be able to easily access this information, we scale the node-size according to the total number of members of each community. The final piece of information is the ratio of $\Delta$-nodes to $\Gamma$-nodes, which we can obtain by coloring the node (e.g. as a pie-chart). Figure~\ref{fig:communitynet} displays a number of such networks of communities for the cond-mat network.

Let us think about the expected behavior of the network of communities. In the case of $K_{1,1}$ (cf.~Figure~\ref{fig:communitynet}~a), the network of communities is simply one large node displaying the fraction of $\Delta$ and $\Gamma$ nodes. When we increase $a$ and $b$ in $K_{a,b}$, this node breaks apart into smaller pieces. If the network is highly modular, the resulting network of communities will be quite sparse and many nodes will have degree zero; if the network is homogeneous, we find a densely interconnected network of communities. For a given choice of $a$ and $b$, the structure of the resulting network of communities provides a useful way estimate of the information content of the individual communities.

The network of communities illustrates what aspects of community structure we are probing, when we adjust the values of $a$ and $b$. This is illustrated in Figure~\ref{fig:communitynet}. Panel~(a) shows the network of communities for $K_{1,2}$. Displayed here is the connected component in the paper-network and the pie-chart shows the fractions of authors and papers in the network.

\begin{figure*}
\centering
\begin{tabular}{c}
\includegraphics[width=\hsize]{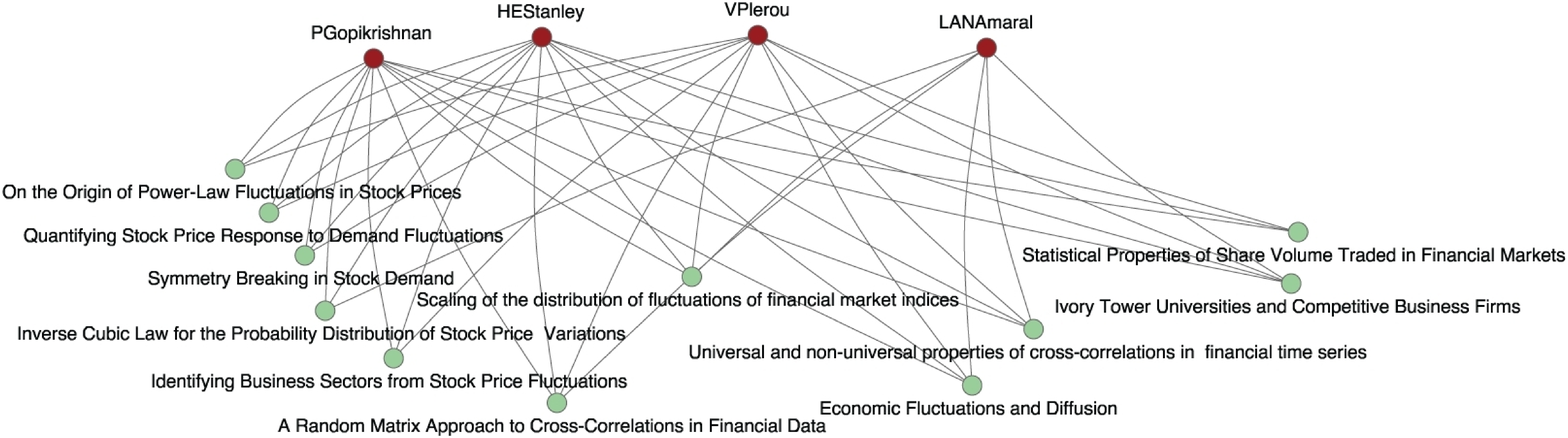} \\
\includegraphics[width=\hsize]{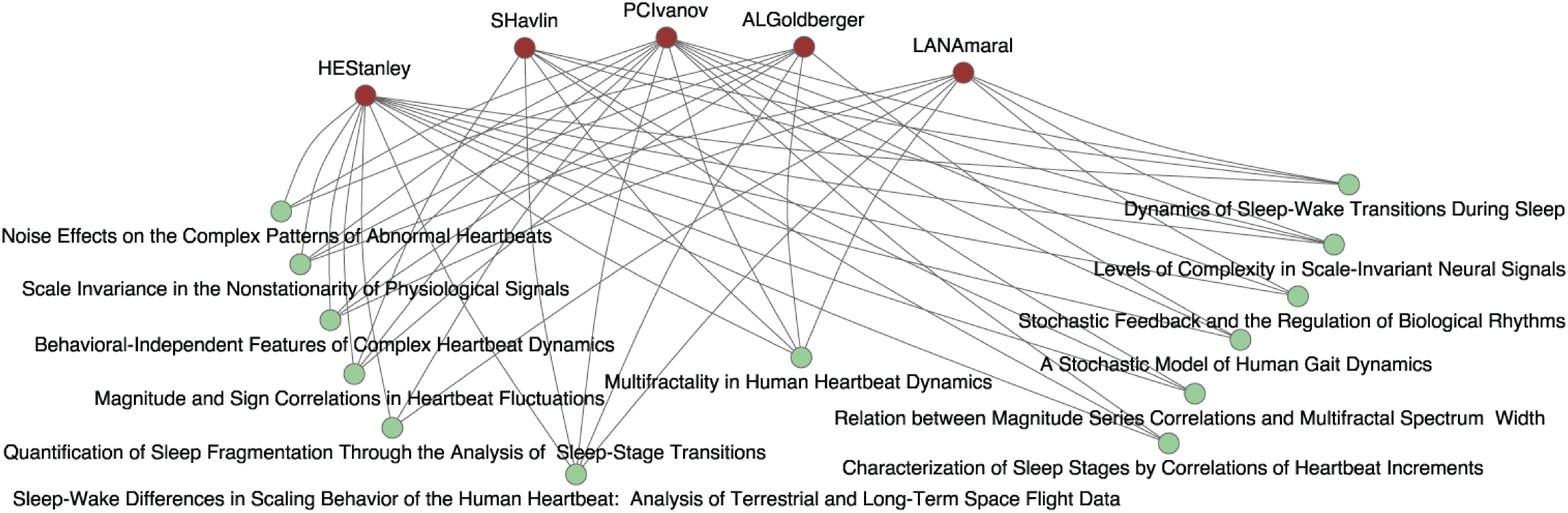}
\end{tabular}
\caption{Color online: The biclique algorithm in action on the cond-mat network \cite{arxiv} (years: 1996-2006). The top panel shows a $K_{3,5}$-clique community of $4$ authors and $11$ papers; this community is a group of scientists studying econo-physics. The bottom panel shows another $K_{3,5}$-clique community, this time consisting of $5$ authors and $13$ papers. The topic of this second community is bio-physics, more specifically analyses of various biological time-series. A key point is that two authors (H. E. Stanley and L. A. N. Amaral) are members of both communities. The division into biclique communities make it immediately underlines the importance of node-overlap: There is no doubt that Stanley and Amaral are full members of both communities. However, it is also immediately clear why the communities are distinct: they regard different subjects. The presence of context (a list of authors are complemented by a list of papers and vice versa) highly enriches our understanding of the communities. A list of authors and papers in these two communities can be found in the Appendix. Both panels are screen-shots from BCFinder \cite{bcfinder}.
}\label{fig:example}
\end{figure*}

Figure\ref{fig:communitynet}~(b) shows the network of communities based on $K_{8,2}$-cliques. The emphasis here is on a large number of shared authors, and as a consequence, each community is dominated by authors. The vast majority of links are dark red indicating author-overlap between the communities. 
Figure\ref{fig:communitynet}~(c) shows the network of communities for $K_{3,5}$. Here the ratio of authors to papers in each community mirrors the global ratio, and all of the communities are of similar in size. In this case the node overlap is equally distributed between author- and paper-overlap. The typical link weight in this network is zero or one. See Figure~\ref{fig:example} for a detailed discussion of two $K_{3,5}$ communities. Finally, panel~(d) shows the network of communities for $K_{2,12}$. In this case, the emphasis is on many shared papers, so all communities contain many more papers than authors (they are mostly light green). Similarly, the majority of links are paper-links; the typical weight is small, between zero and two, but a few heavy links also exist. This threshold probes a completely different aspect of the bipartite network than the $K_{8,2}$ communities.

The networks in Figure~\ref{fig:communitynet} reveal how to analyze the network. If we wish to detect groups of longtime collaborators, we choose small $a$ and large $b$, in this case each community contains only a few authors and many papers, while the overlap with other communities of other longtime collaborators will mainly be papers. The largest community in Figure~\ref{fig:communitynet}~(b) has $12$ authors and $290$ papers, but such a large collaboration is the exception rather than the rule; most communities contain longstanding theoretical collaborations among $2-4$ authors who have written between $20$ and $60$ papers together. If we wish to search for large collaborations, we choose large $a$ and small $b$:  This allows us to find communities of large (typically experimental) collaborations; in this case the communities contain many authors and few papers, while the node-overlap with other communities consists of authors. In the middle interval when $a$ is around the same size as $b$, we find balanced groups of medium size that overlap each other both with papers and authors. If a network is highly modular (as is the case for the cond-mat network), the size of overlap is typically very small, but in dense, more homogeneous networks, the overlaps can constitute a significant fraction of the nodes in each community. The considerations above are specific to the cond-mat network, but a similar analysis can be performed on any bipartite network.

\section{Algorithmic complexity} 
The algorithm proposed above can be used to analyze large sparse networks efficiently. In analogy the problem of enumerating all maximal cliques (which is a classic NP complete problem \cite{karp:72}, which must be solved to detect $k$-clique communities), the problem of enumerating all maximal bicliques is NP complete \cite{peeters:03a}. Roughly speaking, the problem is NP complete because the number of maximal bicliques, $N$, can grow exponentially as a function of the size of the input data. However, as we shall see in the following, this is rarely problem in sparse real world networks. Modern algorithms exist that are very efficient on sparse graphs \cite{liu:06fk,uno:05uq} . The algorithm that we utilize \cite{uno:05uq} has a computational complexity of this step proportional to $N$ in the network being analyzed (with respect to memory usage, this algorithm is also quite efficient---the memory usage scales linearly with the size of input data).

Figure~\ref{fig:complexity} shows how $N$ scales linearly as a function of the number of edges $M$ in two large real world networks: The IMDb network of actors and movies \cite{imdb} and the network of scientific authors publishing in the cond-mat section of the arXiv database \cite{arxiv}. In the case of IMDb, the data for the plots in Figure~\ref{fig:complexity} was created by beginning with the network of all male actors and moves in 1965 and constructing the adjacency matrices and running LCM; then the data for female actors in movies from 1965 was added and the procedure repeated. We expanded the network gradually until it encompassed all movies and all actors and actresses from 1965 to 1980. Separating the male and female actors has the consequence that the number of movies only grows half as often as the number of actors---this accounts for the step-like growth of the black solid line in Figure~\ref{fig:complexity}. In the case of the cond-mat data, a similar method was used, gradually expanding the adjacency matrix 1992, including subsequent years incrementally until 2006. The same procedure is applied to a randomized version of each data set.  Figure~\ref{fig:complexity}~(a) shows the number of maximal bicliques as plotted vs.~the number of edges in the two networks (IMDb, solid black line; cond-mat, solid grey line). The differences between the real and randomized data sets (IMDb randomized, dashed black line; cond-mat, dashed grey line) display clearly that there is significant additional clique structures in the real network data. Figure~\ref{fig:complexity}~(b) shows how the number of nodes grow as a function of the number of edges. In the case of IMDb, the final network contains $497\,386$ edges connecting $163\,416$ actors to $41\,917$ movies. This network contains some $372\,833$ maximal bicliques that it takes the LCM algorithm $3.1$ seconds to locate using a standard lap-top with a 2.16 GHz Intel Core 2 Duo processor and 2 GB RAM. In the case of cond-mat, the final network contains $217\,690$ edges connecting $46\,622$ authors to $70\,975$ papers. This network contains some $81\,697$ maximal bicliques that it takes the LCM algorithm $0.7$ seconds to locate on the same computer.

Creating the overlap matrices and thresholding is $O(N^2) \propto O(M^2)$, where $M$ is the number of edges in the network. Finding connected components in the overlap matrices can be done in $O(N+M_L)$, where $M_L$ is the number of edges of edges in the overlap matrix $L$ and since this matrix is also sparse we have  $O(N)$ for the connected components. These steps are the algorithmic bottleneck; the processing time is a little over 30 minutes for the cond-mat network on the hardware mentioned above. The total complexity of the algorithm is $O(N^2) \propto O(M^2)$\footnote{This result is accompanied by the caveat that the number of bicliques in very dense networks might not scale linearly with the number of nodes.}.  Since the process of finding the bicliques is rather involved, we have created a tool (BCFinder \cite{bcfinder}) that is able to automatically detect and display biclique communities. BCFinder may be freely downloaded.

\section{Discussion}
We have presented a novel method for detecting communities in bipartite networks. Our method is based on an extension of the $k$-clique community detection algorithm suggested by Palla \emph{et al.} \cite{palla:05a}, and explains the relation between the biclique communities and the communities in the corresponding unipartite graph. If bipartite information is available, the algorithm retains all of the advantages of the $k$-clique algorithm (overlapping nodes, the ability to find the network of communities in a given network, etc.), avoids discarding important structural information when projecting the network, and provides a new level of flexibility due to the two thresholding parameters $a$ and $b$, cf. Section~\ref{sec:noc}. The biclique method is computationally manageable for many sparse networks; in cases where the number of bicliques scales linearly with the number of links (as it is the case for the networks analyzed here), the algorimic complexity scales like $O(M^2)$, where $M$ is the number of edges in the bipartite network. 

While our purpose here is mainly to present and analyze a new approach for detecting communities in complex bipartite networks, it is nonetheless instructive to see a small example of the algorithm in action. Figure~\ref{fig:example} shows the algorithm applied to a real network, the cond-mat network of authors and scientific papers from 1996 to medio 2006. The top panel shows a $K_{3,5}$-clique community of $4$ authors and $11$ papers; this community is a group of scientists studying econo-physics. The bottom panel shows another $K_{3,5}$-clique community, this time consisting of $5$ authors and $13$ papers. The topic of this second community is bio-physics, more specifically analyses of various biological time-series. A key point is that two authors (H. E. Stanley and L. A. N. Amaral) are members of both communities. The division into biclique communities make it immediately clear that it is important that communities are allowed to overlap: There is no doubt that Stanley and Amaral are full fledged members of both communities. However, we also understand why the communities are distinct: they regard different subjects. The presence of context (a list of authors are complemented by a list of papers and vice versa) highly enriches our understanding of the communities; this information is not available from the one-mode projections. A list of authors and papers in these two communities can be found in the Appendix. 

We expect that the biclique community detection algorithm will be of practical importance in all areas where the networks studied are bipartite (biological networks, affiliation networks, information networks).
\acknowledgments

This work is supported by the Danish Technical Research Council, through the framework project 'Intelligent Sound', www.intelligentsound.org (STVF No. 26-04-0092) and by the Danish Natural Science Research Council. In addition, the work is supported by the James S. McDonnell Foundation 21st Century Initiative in Studying Complex Systems, the National Science Foundation within the DDDAS (CNS-0540348), ITR (DMR-0426737) and IIS-0513650 programs, as well as by the U.S. Office of Naval Research Award N00014-07-C and the NAP Project sponsored by the National Office for Research and Technology (KCKHA005). 
\appendix
\section{Tables}

\begin{table*}
\begin{tabular}{ll}
\textbf{Authors} & \textbf{Papers}\\\hline
H.E.~Stanley & On the Origin of Power-Law Fluctuations in Stock Prices\\
P.~Gopikrishnan	& Quantifying Stock Price Response to Demand Fluctuations\\
V.~Plerou & Symmetry Breaking in Stock Demand\\
L.A.N.~Amaral & Inverse Cubic Law for the Probability Distribution of Stock Price Variations\\
&	Universal and non-universal properties of cross-correlations in financial time series\\
&	A Random Matrix Approach to Cross-Correlations in Financial Data\\
&	Scaling of the distribution of fluctuations of financial market indices\\
& 	Economic Fluctuations and Diffusion\\
& 	Identifying Business Sectors from Stock Price Fluctuations\\
&	Statistical Properties of Share Volume Traded in Financial Markets\\
&	Ivory Tower Universities and Competitive Business Firms
\end{tabular}
\caption{Community displayed in top panel of Figure~\ref{fig:example}.}
\end{table*}

\begin{table*}
\begin{tabular}{ll}
\textbf{Authors} & \textbf{Papers}\\\hline
S.~Havlin & Scale Invariance in the Nonstationarity of Physiological Signals\\
H.E.~Stanley & Noise Effects on the Complex Patterns of Abnormal Heartbeats\\
P.C.~Ivanov & Behavioral-Independent Features of Complex Heartbeat Dynamics\\
A.L.~Goldberger & Sleep-Wake Differences in Scaling Behavior of the Human Heartbeat:\\
& Analysis of Terrestrial and Long-Term Space Flight Data\\
L.A.N.~Amaral & Magnitude and Sign Correlations in Heartbeat Fluctuations\\
 & Dynamics of Sleep-Wake Transitions During Sleep\\
 & Levels of Complexity in Scale-Invariant Neural Signals\\
 & Relation between Magnitude Series Correlations and Multifractal Spectrum Width\\
 & Multifractality in Human Heartbeat Dynamics\\
 & A Stochastic Model of Human Gait Dynamics\\
 & Stochastic Feedback and the Regulation of Biological Rhythms\\
 & Quantification of Sleep Fragmentation Through the Analysis of Sleep-Stage Transitions\\
 & Characterization of Sleep Stages by Correlations of Heartbeat Increments
\end{tabular}
\caption{Community displayed in bottom panel of Figure~\ref{fig:example}}
\end{table*}

\bibliography{mainbibliography}
\bibliographystyle{unsrt}
\end{document}